**Title: Toward affect-inclusive models of cognitive dynamics: Coupling epistemological resources and emotions**

Authors: Ayush Gupta[1], Brian A. Danielak[2], Andrew Elby[1,2]

[1]*Department of Physics, University of Maryland, College Park, MD 20742.*
[2]*Department of Teaching and Learning, Policy and Leadership, University of Maryland, College Park, MD 20742.*

# Toward affect-inclusive models of cognitive dynamics: Coupling epistemological resources and emotions

Ayush Gupta[1], Brian A. Danielak[2], Andrew Elby[1,2]

[1]*Department of Physics, University of Maryland, College Park, MD 20742.*
[2]*Department of Teaching and Learning, Policy and Leadership, University of Maryland, College Park, MD 20742.*

Many prominent lines of research on student's reasoning and conceptual change within learning sciences and physics education research have not attended to the role of learners' affect or emotions in the dynamics of their conceptual reasoning. This is despite evidence from psychology and cognitive- and neuro- sciences that emotions are deeply integrated with cognition and documented associations in education research between emotions and academic performance. The few studies that have aimed to integrate emotions within models of learners' cognition, have mostly done so at a coarse grain size. In this manuscript, toward the long-term goal of incorporating emotions into fine-grained models of *in-the-moment* cognitive dynamics, we present a case study of Judy, an undergraduate electrical engineering and physics major. We argue that a fine-grained aspect of Judy's affect, her annoyance at a particular kind of homework problem, stabilizes a context-dependent epistemological stance she displays, about an unbridgeable gulf she perceives to exist between real and ideal circuits.

## 1. INTRODUCTION

### A. Research on learner's conceptual development does not attend to the role of learners' emotions in learning

As instructors, educators, and education researchers, intuitively we know that students' emotions, motivations, and sense of identity can strongly influence their behaviors in science learning environments. Research in cognitive science makes a persuasive case for the role that affect plays in regulating performance on cognitive tasks [1–7]. Nonetheless, explicitly incorporating emotion into theories of science learning, though recognized as important, has proved to be a challenge [8–11].

Cognitivist accounts of student reasoning and conceptual change in science have primarily focused on the form and content of novices' knowledge, but haven't attended to learners' emotions [12–17]. Within research on physics learning also, the focus has mostly been on the content of students' ideas [18–24]. Two resource papers on physics education research, assembling important research in the field list research on students' ideas in content areas such as mechanics or thermodynamics, or on practices such as problem-solving [25,26]. Research on physics learners affect and emotions has not garnered enough attention yet to deserve a category mention within these two collections of resources on physics education research and the papers in other categories don't have a meaningful or deep treatment of the relation between affect/emotion and learning.

More recently, research has expanded beyond a focus on students' conceptual knowledge to include attention to students' "personal epistemologies," individuals' beliefs towards knowledge and learning in the discipline [27]. Within physics education, this has been motivated by the desire to cultivate sophisticated attitudes towards science [28–31], as well as by the evidence that students' personal epistemologies influence what conceptual knowledge they draw on when learning and doing physics [32–39]. The latter thread conceptualizes personal epistemologies in terms of fine-grained cognitive resources that codify information about knowledge, knowing, and learning, such as *knowledge as transmitted stuff* as a resource for understanding the source of knowledge as propagated from a source to a recipient, and *facts* as a resource for understanding the form of knowledge as separable bits of information [40,41]. These elements contextually activate in different combinations to constitute an individual's stance towards the nature and form of knowledge they are engaging with in a given moment. Over time, repeated activation of particular combinations could potentially lead to coherent 'belief-like' entities [41,42]. However, this line of research on physics learners' epistemologies also does not include emotions within it's purview (with one exception [33] that we will discuss below).

There has, however, been some progress in understanding that learners' emotions contribute to the learning processes. While there are many threads in this literature, in this brief review we focus on the research that aims to establish that learners' conceptual and epistemological development is tied to their emotions. Given our goal of moving toward affect-inclusive cognitive models, we do not review the research that explores constructs such as "interest," "motivation," or that focuses on learners' emotions without a simultaneous attention to learners' conceptions or epistemologies.

**B. Research suggests that learners conceptual and epistemological reasoning might be tied with their affect**

Even early on, a few researchers did allude to the notion that learning is affective. The notion of cognitive dissonance, for example, is acknowledged to have an affective dimension [43–45]. In a now classic manuscript, Pintrich, Marx, Boyle [9] argued for greater attention to "hot" cognition: models of cognition and learning that attend to the conceptual as well as the affective domains. The Cognitive Reconstruction of Knowledge Model [46,47] and the Cognitive-Affective Model of Conceptual Change [48] are two examples of models that aim to integrate learners' motivations and interests in models of conceptual change. The Cognitive-Affective Model of Conceptual Change also allows for emotions other than motivation, such as anxiety, the feeling of threat, etc. to play a role in the dynamics of conceptual change, though research needs to explore further how these aspects play out *in the moment*. On the empirical side, there is evidence that different emotions might have varying levels and quality of influence on learning. Pekrun et al. [49], for example, document that higher academic achievement is positively correlated with academic joy and excitement; and negatively with affective states such as boredom. In a similar vein, the affective states of flow and curiosity have been shown to be correlated with greater learning gains [50,51].

Some research has attended to the intertwining of epistemological beliefs and affect, though much of this research has been carried out in the context of mathematics education [8,52–57]. Boaler and Greeno [52] found that in a traditional, authoritarian classroom environment a student can develop perceptions of mathematics as an un-creative subject. For some students, this clashes with their personal value toward understanding an idea via exploring multiple interpretations, with the epistemological mismatch producing negative affect toward the subject itself. Op't Eynde et al. [55] take a socio-constructivist approach to emotions arguing that learners' emotions during problem solving are intertwined with their interpretations and appraisals of the unfolding activity, and these in turn are intertwined with their epistemological and self-efficacy beliefs. Cobb et al. [53] have argued that the formation of socio-mathematical norms in the mathematics classroom (some of which have epistemological aspects – norms about what counts as knowing and doing mathematics) are often influenced by the teachers' and students' emotions. Within physics education research, we could only find one article that explicitly argues for the association of students' personal epistemologies and their emotions: Bodin and Winberg [33] found that students' performance on a classical mechanics assignment was associated with expert-like beliefs (such as sense-making) and emotions associated with control and concentration, with the beliefs and emotions being measured through self-reports on survey instruments.

### C. Our goal is to shed light on *how* affect influences learning through building models that couple conceptions, epistemologies, and emotions at fine-timescales

Correlational studies [33,49,50] are significant in that they furnish empirical evidence *that* emotions influence learning, but they do not provide *causal mechanisms* through which the observed correlations arise. Models of affective conceptual change aim to incorporate affect, but are limited by their (1) attention to motivation and interest, (ii) conceptualization of novice knowledge as stable organized structures that transform via learning. Recent empirical evidence, however, points to a view of novice knowledge as more dynamic, unstable, and contextual [34,36,38,58–62]. Researchers have argued for a similar dynamic, contextual view of students' personal epistemology [41] as well as for emotions [55,63]. Given this view of knowledge, beliefs, and emotions as dynamic at fine timescales and the emerging evidence that the emotional and cognitive domains are interconnected [2,3,5,64,65], it seems reasonable to ask: What is the role of students' emotional stances in the dynamics of their conceptual and epistemological reasoning *on a moment-to-moment basis*?

In this paper, we use a case study of an undergraduate engineering major, "Judy," to argue for the feasibility and explanatory power of integrating conceptual, epistemological, and affective elements at fine grain sizes and fine time scales. Our analysis shows that in some contexts, Judy, an electrical engineering and a physics major taking a Basic Circuits course, argues that there is a wide gulf between real circuits and idealized circuit models taught in the course, a gulf that renders qualitative understanding of ideal circuits practically useless to her. We will argue that an affective stance—Judy's deep *annoyance* at qualitative, conceptual homework problems—stabilizes her epistemological stance. Our analysis aims to show how disrupting the affective stance co-occurs with the reversal of the epistemological stance – at least briefly.

This argument has instructional implications. If a learners' emotional stance plays a significant role in sustaining their epistemological stance, as we argue is the case with Judy, then interventions targeting the learners' epistemologies or conceptions directly (for e.g. the intervention described by Rosenberg, Hammer, and Phelan [38] need not be the only or the productive instructional pathway.

In the following section, we present the methods of our case study. We then present the patterns we found in Judy's epistemological and emotional stances. Finally, we make the case that a fine-grained account of Judy's reasoning is empirically inadequate if it does not include the affective elements that lend stability to her epistemological stances. The argument relies on competing toy models of Judy's cognitive dynamics, one of which includes affect and the other of which does not.

## II. DATA COLLECTION AND METHODOLOGY

We videotaped clinical interviews of electrical engineering majors in a basic circuits course at a large public university. The course homework and exams mixed "traditional" quantitative problems with conceptual questions asking students to interpret equations and/or explain physical processes (rather than plugging and chugging). These conceptual problems were co-created by the course instructor and two of the authors (AG, AE). One of us (BD) conducted one-hour semi-structured interviews with Judy and with 3 other students to explore their approaches to mathematics within, and their views and feelings about, the course. Interview participants were paid ten dollars.

First, we started with one of us (BD) viewing the videotapes from the interviews and in real time generating content logs of the interview – brief summaries of the conversation and activities in each 3-5 minute segment. Given our research interest in students' approaches to learning and problem-solving within the course, a few things were specially attended to and flagged at this stage: (i) explicit epistemological statements made by students, (ii) explicit details on their experience in the course, especially if it pertained to the use of mathematics, (iii) meta-conversational features such as large gestures, or explicit facial expressions.

Then we looked at the interview videotapes together as a group. Here mostly the tape is run in real time. And anyone is welcome to stop the tape if they notice anything that they want to flag or talk about [66–68]. Working as a group, we looked for epistemological and affective patterns in participants' responses, both within and across subjects. While not engaging in systematic interaction analysis [69], we informally borrow tools from discourse analysis [70], framing analysis [71], and affective analysis [54,55] to interpret gestures, facial expressions, word choice, and the contextualized substance of participants' utterances in the interviews.

In this paper, we present our analysis of one of our participants, Judy – a second year Electrical Engineering and Physics major. Watching the interview with Judy, very early in the process, what jumped out at us was (i) her emotional reactions when she talks about the conceptual questions in her Basic Circuits homework and (ii) a strong sense of rift she perceived between qualitative understanding of ideal circuits and working with real circuits. We formed explanations for her reasoning and behaviors and then looked for evidence elsewhere in the data to confirm or disconfirm the validity of these patterns [72,73]. Working from a knowledge-in-pieces perspective [15,74], we did not expect global coherence in all aspects of her thinking; we continually considered how specific contextual cues might trigger different "local coherences" in her thinking [15,35,38]. Drawing from notions of "society of mind" [75] and knowledge-in-pieces [15] we model Judy's reasoning in the interview in terms of dynamic activation of epistemological [38,41] and affective "pieces."

## III. DATA AND FINDINGS: NOTICING EPISTEMOLOGICAL AND AFFECTIVE PATTERNS IN JUDY'S TALK ABOUT CIRCUITS

In this section we chronologically present segments of the interview that illustrate that (i) Judy thinks that there is a wide gulf between real circuits and idealized circuit models, (ii) that conceptual reasoning about idealized circuits is unproductive and useless and (iii) Judy is has a strong negative affective reaction (annoyance) towards the conceptual reasoning about circuits.

The interview starts with Brian asking Judy about the course. Judy states that the course is challenging since this is her first Electrical Engineering course while other students had taken a course or two before this. When asked, she says that she thinks that problem-solving is the most important aspect of the course. She acknowledges that she does not know what a professional electrical engineer does but believes that it must involve dealing with circuits. To delve further into her sense of what circuits knowledge a professional electrical engineering might need, Brian asks Judy to imagine a female professional engineer who is taking a basic circuits course to reinforce her understanding of circuits, in preparation for a new job assignment. Specifically, Brian asks what a perfect version of the Basic Circuits course would look like for that engineer:

> [05:16] Judy: Well, I think, um, the course should talk more about, um, how the actual world works. Because, sometimes we talk about, like, ideal circuits and, um (*frowns*) theoretical methods. Those are not related to the actual circuit and those kind of things. So if the professor can talk a little bit more about the actual circuit and how those work, then it may be better for her. (*sighs and relaxes to a less alert posture*)

[05:55] Brian: Why is that difference important to you?

[05:57] Judy: Um, (*smiles slightly*) Because the ideal world is different. (*slight laugh as she utters the last words*)

[06:02] Brian: Ok

[06:13] Judy: Um, and, sometimes [in class] we talk about, um, the physical aspects of the circuits. Um, but I feel like when you really work on circuits it's—it's not very important. I mean, it's better to know what happened physically, but if those are not ideal circuits, then it's different.

Here (and elsewhere), Judy expresses the view that the "real world" of circuits operates much differently from the "ideal circuits" she has learned about in class; the second is not a close approximation of the first. This disconnect, Judy posits, makes learning how actual circuits work more valuable for the hypothetical professional engineer.

The real-ideal distinction holds personal importance for Judy. As Judy talks about "ideal circuits and theoretical methods" she frowns and her facial expression registers displeasure. As Judy speaks about the lack of utility in understanding the "physical aspects of the circuits", she frowns again and her face pulls to a side in an expression of mild disgust. It is difficult to place a label to her emotion, but not difficult to understand the negative valence of her affect in this moment. The emphasis in Judy's speech as she utters "really work on circuits" underscores the difference she sees between working on circuits in real life and the problems and discussions that are a part of the basic circuits course. This real-ideal distinction, for her, is coupled to another pattern that comes up often in the interview: that learning about the idealized and/or physical models of circuits does not have practical use.

[06:45] Brian: Hmm (*pause*) So do you think if you're analyzing a real-world circuit, it's still important to know about the physical aspects of the circuit?

[06:55] Judy: Not very important

[06:57] Brian: OK. What is important?

[07:01] Judy: What is important?

[07:02] Brian: Mmmhmm. Especially if you're dealing with a real world circuit.

[07:06] Judy: Know exactly how the- what's the difference between a real world circuit and an ideal world circuit, and, yeah how to deal with the real one

Brian tries to probe a bit deeper into this real-ideal gulf that Judy feels and the associated sense that conceptual reasoning about ideal circuits is not useful for analyzing real circuits.

[08:00] Brian: When you started the course, did you have a sense of differences between ideal world circuits and real world circuits?

[08:09] Judy: No, I know nothing about circuits.

[08:12] Brian: OK.

[08:17] Brian: So, for you as you've taken the course, has your sense of real-world and ideal world changed in any way since the beginning?

[08:28] Judy: Yeah. Yes, I now feel like many components in the circuits are not perfect. It's not like you can completely use the physical method to analyze those.

[08:46] Brian: OK

[08:50] Brian: Could you say a little bit more about the physical method?

[08:53] Judy: Um, like a capacitor, I might say this wrong, but um, like they have a limit that how many currents, or how many electrons (*moves left hand up and down in a chopping motion*) can stay on the plate (*chops both hands down*). That's a real world (*moves left hand up and down in a chopping motion*) capacitor. But, for, ideally, sometimes in the course we assume that each plate (*spreads hands apart*) can have infinite amount of electrons (*turns both hands palm up*), and those kind of things. So, I mean, if that doesn't exist (*turns left hand palm up*) then why do we use that method to solve it? We are not gonna use that method in the future anyway. I mean, it's good to know, but it's...yeah.

Judy's sense that real circuits components behave differently from the idealized models could be a healthy skepticism; many expert engineers will also hold that view. But for experts that skepticism is typically balanced with a view that the models are productive in generating baseline information about how a real system will behave and in the design of real circuits. Even when idealized models do not represent reality, they can be useful in generating solutions to real life issues [76]. But the stance Judy takes is that these methods of conceptual reasoning – because of their distance from the real world – will be completely useless when someone starts working on a real circuit.

Judy uses the language of ideal world, physical aspects, and theoretical methods fairly interchangeably. Judy's references to "physical aspects" seemed to mirror some of the prompts in the revised assignments for the course, where students were often asked to present a physical (non-mathematical) argument for how a circuit will behave. It was sometimes explicitly specified that the physical argument is conceptual or does not involve carrying out the mathematical manipulations of equations. In most cases, questions that asked for "physical reasoning" were followed by those requiring calculations and looking for consistency between the conceptual argument and results of numerical or symbolic computations. As the interview proceeds, Judy, like other students we interviewed, distinguishes between the regular, mathematical problems and the more conceptual problems on the homework:

[15:33] Brian: A student whose opinion I heard earlier from your class had noticed that the homework and tests seemed to contain two sort of types of questions—two different kinds of questions. And I was wondering if it's been your experience that you've noticed something like that.

[16:01] Judy: Oh. I think, um, you mean two parts, right? Umm, I think one of them is like, um, just problem-solving. Like, you have a diagram (*spreads hands out, gently taps table with fingertips*) and then you solve for the current or voltage. Um, and the other type is like (*shakes head gently side-to-side*) a physical question. They will ask you what is (*shakes head*) physically (*pinches thumb and forefingers of both hands together, moving her hands up and down*) happening in the circuit, and you have to explain them in (*shakes head slightly, left to right*) words.

[16:31] Brian: OK.

[16:33] Judy: Um, because, um, I am also double-majoring. My second major is physics, so I guess it's good to know more about like, what's physically happening (*pulls her mouth to side*) in a circuit. Like for me, cause I'm--that's my major. But, if, um--for a student who's just doing an EE major, I don't think that's very necessary.

Just as she distinguishes between real world and idealized models, Judy distinguishes between course assignment questions focused on conceptual reasoning and those focused on mathematical manipulations ("problem-solving"). She ties the conceptual reasoning questions with idealized models, and speaks of their limited practical use for an engineering student (as seen above), while highly valuing the role of problem-solving questions (we discuss evidence for this later). Her stance towards the value of conceptual reasoning is nuanced though; she thinks that understanding mechanistic models underlying the phenomenon ("what's physically happening in a circuit) is valuable for a physics student. So it's possible that the value divide that she sees derives from some deeper epistemological discontinuity that she sees in physics and engineering.

Though our attempts to transcribe gestures cannot fully capture the annoyance Judy displays when discussing the "physical question[s]," it was strongly present here and elsewhere in the interview. Figure 1 shows a snapshot of her expression at time-stamp 16:45, right at the moment when she is uttering "physically happening" in the transcript above.

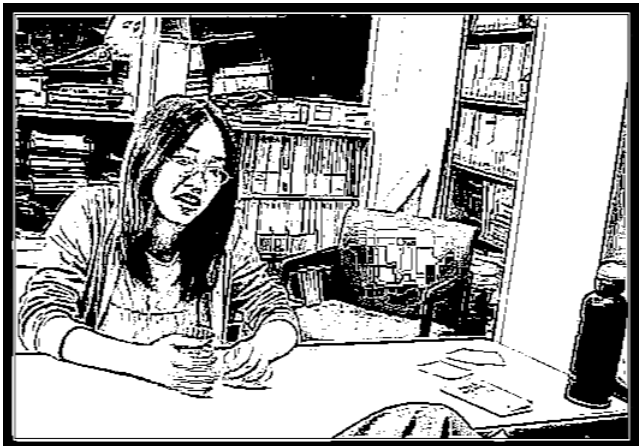

**Figure 1 Judy's expression showing negative affect when talking about what's "physically happening" in the circuit (timestamp 16:45)**

Finishing her last utterance (16:33), Judy explicitly labels these questions as "annoying":

[17:13] Judy: Yeah. I mean those questions are—it's (*shakes head gently, but repeatedly*) kind of annoying (*smiles*). But it's good to know (*rise in voice tone*).

[17:24] Brian: What is it that makes them annoying?

[17:25] Judy: It's hard to under--to understand or to answer. It's, umm. Yeah maybe we don't know enough about physical things. Because, in class we don't talk about those. We don't--in class we just do the problem-solving part. So, um, yeah. It's hard to answer. I know many people, um go to the professor's office hours just asking those physical questions.

[18:10] Brian: Have you talked directly with anybody who has a hard time also answering the physical kind of questions?

[18:16] Judy: Umm, yeah, but we didn't talk much.

[18:22] Brian: Hmm, OK.

[18:23] Judy: Uh, yeah. know they feel annoying too.

[18:29] Brian: OK.

Judy's annoyance at the conceptual questions seems to stem, in part, from the lack of instructional support that she perceives for answering those questions, especially since she also thinks of these questions as difficult to answer. And Judy, here, recruits support for her annoyance through claiming that this emotion is shared by her classmates.

Physiologically, we again note that Judy's facial expressions reflect mild displeasure, though at times, she is also smiling, and at times reflecting embarrassment. The substance of her utterances, however, here provide greater evidence for her annoyance. Here and elsewhere, even when we note displeasure, her expressions are muted; and that is expected. The interview is not typically a context for exaggerated expressions. Judy, overall, seems shy during the encounter and speaks in a soft voice. Most importantly, emotional expressions are not simple reflection of what emotion one experiences, but are also tools for communication in social interaction. So, given the interview context it is not surprising that Judy's expressions are muted. Also, in this particular instance, Judy is speaking about her annoyance rather than being in a scenario where she is experiencing it. All of these caveats bring greater comfort to arguing that Judy does feel annoyed at conceptual reasoning.

Following this exchange, Brian turns again to explore the relevance of the conceptual questions for her projected professional life:

> [18:32] Interviewer: So, um, this is a--this is a sort of difficult question, but, um, with our imaginary person Diana, who's the professional engineer. Do you think she would probably find that those questions are tough also?
> 
> [18:47] Judy: Yes, I guess.
> 
> [18:48] Interviewer: OK. Why do you say that?
> 
> [18:50] Judy: Because [shakes head gently, but repeatedly, left-to-right] those physical questions are not very related to the actual world; not related [knits brow] to her job. So...that's why I say it's not very necessary for student[s] who are only in EE major to learn those parts.
> 
> [19:22] Interviewer: Do you think one kind of question is more helpful to you than the other kind?
> 
> [19:28] Judy: Umm, yeah I think the [cocks head, smiles] problem solving is more helpful [laughs] and [nods] more important.

Thus, Judy feels those questions have little bearing on what engineers actually do. Judy sees conceptual reasoning as useless for practical purposes and considers equation-based problem-solving as "more helpful" and "more important."

We turn now to another characteristic pattern we observed in Judy's behavior. When talking about or engaging in equation-based problem solving, Judy displays positive affect: she smiles and laughs and does not appear to be tensed. Figure 2 shows her general demeanor during answering quantitative problems.

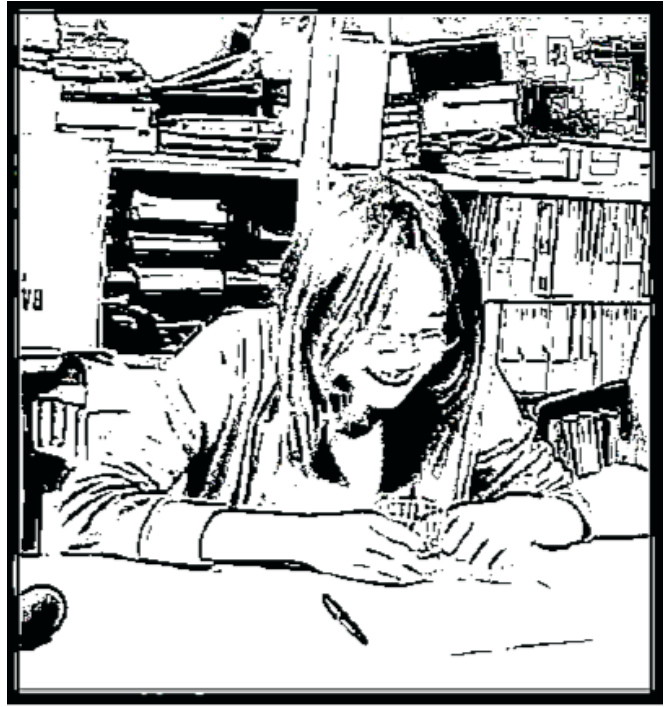

**Figure 2 Judy shows positive affect when working on traditional circuit problems.**

We find no evidence that in these moments she is ever thinking about the real/ideal gulf. This is striking given that these problems refer the same ideal circuits she disparages elsewhere in the interview (an example problem presented to Judy later in the interview is shown in Figure 3). Earlier in the interview, though, it was Judy—not Brian—who brought up the real/ideal gulf when discussing conceptual questions. Her epistemological view about the real/ideal gulf is thus context-dependent: *present* when she discusses conceptual problems, but *absent* in both her discussions of and solutions to traditional ones.

Over the course of the interview we thus noticed repeated affective and epistemological reactions from Judy toward aspects of her Basic Circuits class. We summarize these as four *patterns* – affective and epistemological – that we subsequently use to create cognitive toy models for describing the conceptual, epistemological, and affective dynamics of Judy's behavior during the interview:

*Pattern 1.* Judy was annoyed by the conceptual questions on homework

*Pattern 2.* Judy viewed conceptual reasoning as unproductive to her learning

*Pattern 3.* Judy expressed a gulf between real-world circuits and the ideal circuits probed by the conceptual questions

*Pattern 4.* When discussing conventional quantitative problems, Judy displayed no annoyance and did not discuss the real/ideal gulf from *Pattern #3*

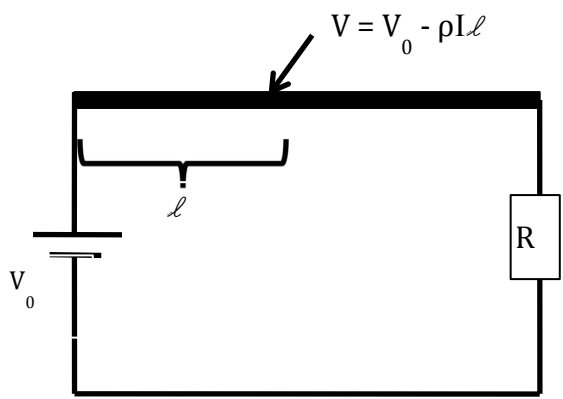

**Figure 3 Example problem from the interview: The potential at a point a distance $\ell$ along the "fat resistive" wire is given by $V = (V_0 - \rho I \ell)$. How will you explain the equation to a friend from your circuits' class.**

## IV. MODELING JUDY'S REASONING AND BEHAVIOR: AFFECT MEDIATED EPISTEMOLOGICAL DYNAMICS

In this section, we model the patterns in Judy's reasoning in the interview in terms of fine-grained cognitive elements. First, we first present two possible "toy cognitive models" [38]: one that includes her epistemological stances but treats Judy's negative affect toward the conceptual questions as epiphenomenal and as such not a part of the model; the second includes her affective state in the model as coupled with the epistemological cognitive elements. Next we trace the implications of these two models for hypothetical situations in which the two models will predict different patterns of reasoning/behavior. And we present two episodes from the interview that align better with the predictions of the affect-inclusive model for Judy.

### A. Two "toy" cognitive models for Judy: Affect-Free and Affect-Inclusive

The patterns above are consistent with a "toy cognitive model" (Rosenberg et al.) in which Judy's epistemological view about the gulf between real and ideal circuits and her sense that the conceptual problems are useless mutually reinforce one another:

> *Real/Ideal Gulf:* an epistemological stance that knowledge relevant to ideal circuits and that knowledge relevant to real circuits are completely disconnected.
>
> *Conceptual Reasoning Useless:* epistemological stance that conceptual reasoning is unproductive, but equation-based problem solving is useful

The cognitive elements we attribute to Judy are context-dependent and dynamically activated rather than being global belief-like. For Pattern #2 and #3 in the context of thinking about the conceptual problems, both the posited elements seem to be active; Judy takes the stance that ideal circuits are so different from real ones as to render pointless thinking about ideal circuits. In pattern #4, by contrast, when she's focused on mathematical problem-solving the *issue* of a possible real/ideal gulf isn't part of her thinking, suggesting that the posited elements of the epistemological coherence are 'off.' (Being 'off' does not imply that she is taking an opposing stance; simply that these elements don't seem to be playing a role in her reasoning and behavior.) Note that this is a toy model, leaving out the vast majority of cognitive elements needed to model her cognition; but it serves the purpose of explaining the patterns of Judy's epistemological stances. Further, the elements that we posit, could also be described in terms of finer grained elements (Real/Ideal Gulf, for e.g. could be thought of as resulting from ontological resources about real world versus ideal constructs, an awareness of two kinds of knowledge related to the two ontologies, and a feeling of strong disconnect between the two kinds.) But for our purpose in this manuscript, the grain-size we choose suffices.

So far, our model does not include Judy's emotions. One possible way of attending to Judy's annoyance is to think of her affect as incidental – Judy is annoyed in the context of conceptual questions, and shows positive affect in equation-based problem-solving contexts. By this account, affect is a by-product of the context but does not stabilize Judy's epistemological stances; as such, it would not be an element of her local cognitive coherence. A second possibility is to include an additional affective element in the toy model of her local cognitive coherence:

> *Annoyance*: negative affective response towards conceptual and qualitative reasoning and questions

Using Judy as an example, we want to distinguish between these two models – one in which affect is part of the local coherence and one in which it isn't – and argue that in some cases, like Judy's, affect plays a role in driving cognitive dynamics and is not just along for the ride. These models can lead to very different empirical predictions that can be used to test which model has more explanatory power. For example, consider an instance where some string of events leads Judy to think conceptually in the context of the "physical aspect" problems that she dislikes. The affect-inclusive model would predict that if her *annoyance* is disrupted in this instance, then her stance that conceptual reasoning is useless would also weaken in this instance. By contrast, the affect-free model would not predict any affective influence on the epistemological dynamics; disrupting her annoyance should not affect her "cold cognitive" epistemological stance.

Here's another way to distinguish between the affect-free and affect-based toy models. Suppose Judy is presented with a real circuit set-up (with no idealizing assumptions made) and is presented with some conceptual reasoning that correctly applies to that circuit. According to the affect-free model, the focus on a non-idealized circuit should suppress (render irrelevant) the stance that real and ideal circuits are distinct (*Real/Ideal Gulf*); and therefore, the *Conceptual Reasoning Useless* stance should be weakened as well. Put more simply, if Judy finds idealized conceptual reasoning useless partly *because* she perceives a gulf between ideal and real circuits, then focusing on real circuits and on conceptual reasoning that definitely applies to that real circuit should make Judy less prone to see conceptual reasoning as useless in this instance. By contrast, our affect-inclusive toy model predicts that, even if *Real/Ideal Gulf* is suppressed, Judy's annoyance at conceptual problems still reinforces her *Conceptual Reasoning Useless* stance. Thus, microanalysis of Judy's local epistemological stances in different contexts could provide evidence for or against the idea that affect plays an integral role in her cognitive-epistemological dynamics.

We scanned the interview for episodes that could provide confirmatory or conflicting evidence for either model at the expense of the other. We found two such episodes. They provide inconclusive but suggestive evidence in support of the affect-inclusive model and against the affect-free one.

**B. Annoyance as part of local cognitive coherence: Episode A:**

At the close of the interview, Brian asks Judy one of the conceptual homework questions from her course.

> [50:06] Brian: Suppose you were comparing, um, three different types of waves in a circuit. So, you're comparing a sine wave, a square wave, and a triangle wave. And all three of them have the same RMS voltage. Which one has the highest peak voltage?
> [50:36] Judy: {*sighs, smiles*} Um, triangle wave.
> [50:39] Brian: OK. How do you know?

Judy talks about how she had found this question particularly challenging, forcing her to turn to her physics TA for help. She summarizes what she remembered of her TA's explanation, acknowledging that she can't remember it fully, and then turns to an equation. She talks about how the answer has something to do with the integral, which is "related to" the area under the curve. Brian then presses her to think it through for herself. In response, she does some excellent informal reasoning:

> [52:58] Brian: So, if you didn't have your book, but you know there was some relationship between the integral and the area. Is there a way that you could think through this?
> [53:07] Judy: Um. {*6 second pause*} I guess—I mean I can try. So they have the same period like from, uh zero to T. So the integral would be like from zero to t. Um, and then they have some kind of wave form here. And they have a same answer. So. {*5 second pause*} I dunno. I mean, if they have the same area, um and, this part is like the same length, um, because like. This looks like wider {*spreads hands apart, past her shoulders*}—like the area looks wider {*spreads hands again*}. So there should be, like. If you squeeze it into {*presses palms toward each other*} this form, it should be like, going up {*forms a peak with her hands*}. So, it should be like this. So, apparently, like, they have a higher peak.

Judy here is thinking of the area under the curve as a kind of squeezable stuff; since the area under the three curves is the same, the amount of stuff is the same, and when you "squeeze" a rectangle (square wave) into a triangle (triangle wave) of the same base length, the triangle ends up with a "higher peak." The Brian probes her stance towards her qualitative, conceptual reasoning:

> [54:42] Brian: But I'm interested in that you said "squeezing." Do you ever, um, ever um think about the mathematics that you use in that way?
> [54:53] Judy:{*Judy laughs*} No, I never think of that before. That's the way how the—my TA told me. Yeah, I mean it's I feel like it's not very formal {*smiles*}, but it's very useful.

Our transcript does not capture the satisfaction—evident in the video—she felt about her solution. Figure 4 shows a still of here facial expressions at this moment.

Though Judy says of her solution, "it's not very formal"—it's the kind of qualitative reasoning she earlier labeled unproductive—in this moment, feeling good about her solution, she calls this kind of conceptual reasoning "very useful." The affect-free toy model doesn't predict such a reversal of her epistemological stance: if *annoyance* is not an integral part of the local coherence, then it is difficult to imagine what would change her view about the uselessness of conceptual reasoning in this moment.

Of course, our argument here assumes that Judy would classify this homework problem, and her reasoning about it, as the kind of "conceptual" question and thinking she disparaged earlier in the interview. She might not, because here she does conceptual reasoning about mathematical objects and operations (integrals, areas under curves) rather than conceptual reasoning directly about circuits. However, small pieces of evidence suggest otherwise. Judy's expressions and posture are tense as the problem is posed; she acknowledges immediately that she had found this question to be very difficult, and she approached her physics TA, not her Circuits course TA, for help; at the end of her successful reasoning, she still tries to distance herself from the "informal" reasoning. Though we cannot make a conclusive case, this evidence suggests that Judy classifies the waveform question and her informal reasoning as the kind of conceptual question and reasoning she calls useless and inapplicable earlier in the interview.

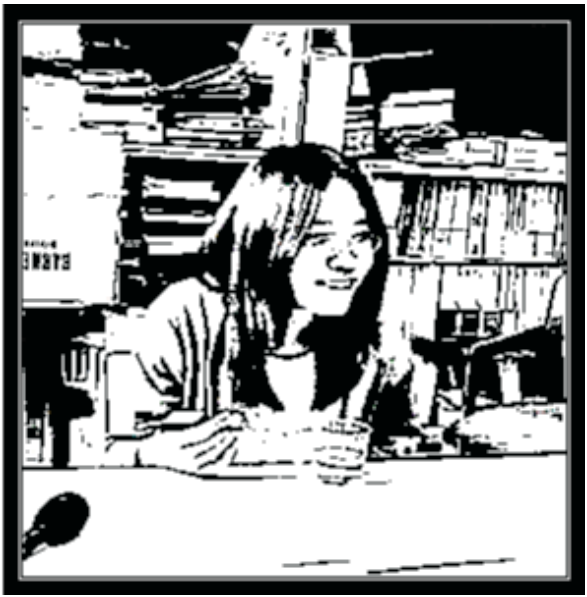

**Figure 4 Judy's positive affect after solving the conceptual problem**

### C. Annoyance as part of local cognitive coherence: Episode B:

At one point (minute marked [06.45] above), while Judy was still thinking about her annoyance at the conceptual problems, Brian probed her views about *both* the real/ideal gulf *and* conceptual reasoning: "So do you think if you're analyzing a real-world circuit, it's important to know about the physical aspects of the circuit?" Judy responds, "Not very important." Crucially, her annoyance at the conceptual problems and the associated conceptual reasoning are still apparent during this part of the interview. As discussed above, a model in which affect is just incidental cannot explain Judy's response: if she finds conceptual problems unproductive partly *because* they apply only to an idealized world, then why would conceptual reasoning be "not very important" when applied to a real-world circuit?

Of course, an affect-free model different from ours might be able to explain Judy's reasoning patterns throughout the interview. But given (i) the previous literature showing the role of affect on epistemological stances, and (ii) the in-your-face nature of the annoyance Judy displays towards conceptual problems during most of the interview, we have reason to take seriously models in which this affective state plays a role in stabilizing or destabilizing Judy's epistemological stances. Incorporating an affective state (*Annoyance*) better explains Judy's behaviors and reasoning:

- *Annoyance* reinforces *Conceptual Reasoning Useless* and the *Real/Ideal Gulf* (Patterns#2 and #3),
- When she perceives herself as not engaged in conceptual reasoning but problem solving, *Annoyance* disappears and so does *Real/Ideal Gulf* in those moments (even though the equations she is using encode idealizations) (Pattern #4).
- Suppressing *Annoyance* can suppress the activation of *Conceptual Reasoning Useless* leading to the reversal of her epistemological stance (Episode A), and

- Targeting Judy's epistemology (by focusing on *real* circuits) without addressing Judy's affect fails to suppress Judy's view that conceptual reasoning is "not very important" (Episode B).

## V. Summary and Implications

At first glance, Judy's patterns of thought (discussed above) seem attributable in part to a robust belief that ideal circuits (addressed by conceptual problems on the homework) are very different from real circuits. We argued, however, that this "belief" is not a robust part of Judy's reasoning across all contexts probed in the interview. Rather, an affective state— annoyance— stabilizes and helps to drive her views about the course's conceptual problems and her epistemological stance about a real/ideal gulf. Including a context-dependent affective state (*Annoyance*) in our model enabled us to explain: (i) the minimal or nonexistent role her epistemological stance about the *Real/Ideal Gulf* plays in her thinking when she discusses or performs mathematical problem-solving, and (ii) the disruption of her stance toward conceptual reasoning induced by an affectively positive experience. In this way, we illustrated how including affect in models of cognitive dynamics provides additional explanatory power.

This analysis has instructional implications. Given Judy's stated belief that the conceptual problems are annoying because of the real/ideal gulf, an obvious instructional strategy might be to directly address the real/ideal gulf by showing how idealizations approximate the real world, stressing the near-agreement of ideal predictions with real results. While we do not disagree with that strategy, we don't think it would be enough for students like Judy. Bringing about epistemological change for Judy (and students like her) might have more to do with setting her up to have affectively positive experiences, and less to do with eliciting, confronting, and replacing her problematic epistemological "beliefs." Ultimately, though limited to one student in one interview, our analysis illustrates the importance and feasibility of incorporating affect into fine-grained models of cognitive dynamics and how doing so can expand the toolbox of productive instructional responses.

## ACKNOWLEDGEMENTS

We thank "Judy" for her participation in the study. We thank the instructor of the Basic Circuits course for allowing us access to students, and incorporating suggested modifications into the curriculum. We thank Eric Kuo, Michael Hull, David Hammer, Jim Pellegrino, Ann R. Edwards, and members of the Physics Education Research Group at the University of Maryland for productive discussions. This work was supported in part by NSF EEC-0835880 and NSF DRL-0733613. The opinions in this manuscript are of the authors only.

[48] M. Gregoire, Educational Psychology Review **15**, 147 (2003).

[49] R. Pekrun, T. Goetz, W. Titz, and R. P. Perry, Educational Psychologist **37**, 91 (2002).

[50] S. D. Craig, A. C. Graesser, J. Sullins, and B. Gholson, Journal of Educational Media **29**, 241 (2004).

[51] S. D'Mello, R. W. Picard, and A. Graesser, IEEE Intelligent Systems **22**, 53 (2007).

[52] J. Boaler and J. G. Greeno, Multiple Perspectives on Mathematics Teaching and Learning 171 (2000).

[53] P. Cobb, E. Yackel, and T. Wood, in *Affect and Mathematical Problem Solving*, edited by D. B. McLeod and V. M. Adams (Springer New York, 1989), pp. 117–148.

[54] V. A. DeBellis and G. A. Goldin, Educ Stud Math **63**, 131 (2006).

[55] P. Eynde, E. Corte, and L. Verschaffel, Educational Studies in Mathematics **63**, 193 (2006).

[56] K. Weber, Research in Mathematics Education **10**, 71 (2008).

[57] M. Zembylas, Journal of Research in Science Teaching **41**, 693 (2004).

[58] D. B. Clark, C. M. D'Angelo, and S. P. Schleigh, Journal of the Learning Sciences **20**, 207 (2011).

[59] D. B. Clark, Cognition and Instruction **24**, 467 (2006).

[60] O. Parnafes, Journal of the Learning Sciences **16**, 415 (2007).

[61] B. L. Sherin, M. Krakowski, and V. R. Lee, Journal of Research in Science Teaching **49**, 166 (2012).

[62] D. Stamovlasis, G. Papageorgiou, and G. Tsitsipis, Chem. Educ. Res. Pract. (2013).

[63] L. F. Barrett, B. Mesquita, K. N. Ochsner, and J. J. Gross, Annu Rev Psychol **58**, 373 (2007).

[64] P. B. Carr and C. M. Steele, Journal of Experimental Social Psychology **45**, 853 (2009).

[65] J. E. LeDoux, Annu. Rev. Neurosci. **23**, 155 (2000).

[66] S. J. Derry, R. D. Pea, B. Barron, R. A. Engle, F. Erickson, R. Goldman, R. Hall, T. Koschmann, J. L. Lemke, and M. G. Sherin, Journal of the Learning Sciences **19**, 3 (2010).

[67] R. E. Scherr, Phys. Rev. ST Phys. Educ. Res. **5**, 020106 (2009).

[68] A. H. Schoenfeld, Journal of the Learning Sciences **2**, 179 (1992).

[69] B. Jordan and A. Henderson, The Journal of the Learning Sciences **4**, 39 (1995).

[70] J. P. Gee, *An Introduction to Discourse Analysis: Theory and Method* (Routledge, 1999).

[71] D. Tannen, *Framing in Discourse* (Oxford University Press, New York, 1993).

[72] B. G. Glaser, *Basics of Grounded Theory Analysis: Emergence Vs Forcing* (Sociology Press Mill Valley, CA, 1992).

[73] M. B. Miles and A. M. Huberman, *Qualitative Data Analysis: A Sourcebook of New Methods.* (Sage Publications, Beverly Hills, CA, 1984).

[74] J. P. Smith III, A. A. diSessa, and J. Roschelle, Journal of the Learning Sciences **3**, 115 (1994).

[75] M. Minsky, *The Society of Mind* (Simon & Schuster, New York, NY, 1988).

[76] J. Gainsburg, Mathematical Thinking and Learning **8**, 3 (2006).